\documentclass[]{aa}

\usepackage{color}
\usepackage{txfonts}
\usepackage{graphicx} 
\usepackage{natbib} 
\usepackage{longtable} 
\usepackage{amssymb}
\usepackage{lscape}
\usepackage{rotating}
\usepackage[english]{babel}
\bibpunct{(}{)}{;}{a}{}{,}

\begin{document}

\title{Revisiting the luminosity function of single halo white dwarfs}

\author{Ruxandra Cojocaru\inst{1,2},
        Santiago Torres\inst{1,2}, 
        Leandro G. Althaus\inst{3,4},
        Jordi Isern\inst{1,2}
        \and
        Enrique Garc\'{\i}a--Berro\inst{1,2}}

\institute{Departament de F\'\i sica Aplicada,
           Universitat Polit\`ecnica de Catalunya,
           c/Esteve Terrades 5, 
           08860 Castelldefels,
           Spain
           \and       
           Institute for Space  Studies of Catalonia,
           c/Gran Capit\`a 2--4, Edif. Nexus 201,   
           08034  Barcelona,
           Spain
           \and
           Facultad de Ciencias Astron\'omicas y Geof\'\i sicas, 
           Universidad Nacional de La Plata,
           Paseo del Bosque s/n, 
           1900 La Plata, 
           Argentina
           \and
           Instituto de Astrof\'\i sica de La Plata, UNLP-CONICET,
           Paseo del Bosque s/n, 
           1900 La Plata, 
           Argentina
           \and
           Institut de Ci\`encies de l'Espai (CSIC), 
           Campus UAB, Facultat de Ci\`encies, Torre C-5, 
           08193 Bellaterra,
           Spain}

\titlerunning{The luminosity function of halo white dwarfs}
\authorrunning{Cojocaru et al.}

\date{\today}

\abstract{White dwarfs are  the fossils left by the  evolution of low-
           and   intermediate-mass   stars,   and   have   very   long
           evolutionary  timescales. This  allows  us to  use them  to
           explore  the  properties  of   old  populations,  like  the
           Galactic halo.}
         {We present  a population  synthesis study of  the luminosity
           function of halo white dwarfs, aimed at investigating which
           information  can be  derived from  the currently  available
           observed data.}
         {We employ  an up-to-date population synthesis  code based on
           Monte Carlo  techniques, that incorporates the  most recent
           and reliable  cooling sequences for metal  poor progenitors
           as  well  as  an  accurate modeling  of  the  observational
           biases.}
         {We  find that  because  the observed  sample  of halo  white
           dwarfs is  restricted to the  brightest stars only  the hot
           branch of the  white dwarf luminosity function  can be used
           for such  purposes, and that  its shape function  is almost
           insensitive to  the most relevant inputs,  like the adopted
           cooling sequences,  the initial mass function,  the density
           profile of the stellar spheroid, or the adopted fraction of
           unresolved  binaries. Moreover,  since the  cut-off of  the
           observed luminosity has not  been yet determined only lower
           limits to the age of the halo population can be placed.}
         {We conclude  that the  current observed  sample of  the halo
           white  dwarf  population  is  still  too  small  to  obtain
           definite conclusions  about the  properties of  the stellar
           halo,  and  the  recently   computed  white  dwarf  cooling
           sequences  which  incorporate   residual  hydrogen  burning
           should be assessed using metal-poor globular clusters.}
 
          
\maketitle
   
\section{Introduction}

White dwarfs are the evolutionary remnant of stars of intermediate and
low masses at the zero-age main  sequence.  The upper limit for a main
sequence star to evolve  to a white dwarf is still  the matter of some
debate, but it is estimated to be $\sim 10\, M_{\sun}$ \citep{Becker1,
Becker2,      Miyaji1980,      Renzini1981,     Nomoto1984,      GBRI,
Poelarends2008}. Thus, given the shape of the initial mass function it
is expected that the vast majority of the remnants of the evolution of
single stars will  be white dwarfs.  Since white  dwarfs are numerous,
have  well  studied  properties  \citep{Althaus2010},  and  have  long
evolutionary timescales, they are the  most suitable tool to study the
properties  of old  populations, like  the Galactic  stellar spheroid.
Moreover, our  knowledge of the  physics controlling the  evolution of
white dwarfs  relies on  solid grounds, since  the basic  principle of
their evolution  is a  well understood  and relatively  simple cooling
process. Although  this basic principle  of the theory of  white dwarf
cooling has  remained unaltered during  the last decades, we  now have
very sophisticated and accurate stellar evolutionary models that allow
us  to  perform  precise  cosmochronology,  and  to  characterize  the
ensemble properties of several white  dwarf populations, like those of
the Galactic disk --  see \cite{Cojocaru2014}, and references therein,
for a  recent work on  this subject -- and  of the system  of Galactic
open   \citep{Garcia-Berro2010,  Bellini,   Bedin2010}  and   globular
clusters \citep{Hansen_2002, Hansen_2013, Gar_etal_2014}.

The population of  white dwarfs in the Galactic stellar  halo has been
the subject of an increased interest since the first observational and
theoretical studies \citep{Mochkovitch90,  Liebert1989}.  Perhaps, one
of the most important reasons of this interest in halo white dwarfs is
their possible contribution  to the dark matter content  of our Galaxy
-- see, for  instance, \cite{Oppenheimer}  for an  observational work,
and \cite{Torres2002}  for a theoretical  study.  However, due  to the
very low space densities and the intrinsic faintness of the population
of white dwarfs  of the Galactic spheroid, their  detection has proven
to be a  difficult endeavour.  Moreover, opposite to  what occurs with
main  sequence  stars  which  can be  classified  according  to  their
metallicity, the atmospheres  of white dwarfs are  devoided of metals.
This  is due  to their  high surface  gravities and  long evolutionary
timescales, which  allow gravitational diffusion to  be very efficient
at  settling  the  metals  resulting from  the  previous  evolutionary
history at the  base of the partially degenerate  envelope.  All these
physical processes make halo  white dwarfs indistinguishable from disk
ones.  Hence,  the only  observational method  to detect  white dwarfs
belonging to the Galactic spheroid  not hampered by relevant technical
difficulties relies on  identifying them on the basis  of large proper
motions,  as radial  velocities cannot  be determined  accurately due,
again, to the large surface gravities, which translates into a sizable
gravitational  redshift  of  the  spectral  features  that  cannot  be
neglected and is  difficult to measure.  Additionally,  the absence of
spectral lines at the very low  luminosities of the coldest halo white
dwarfs  also prevents  an  accurate characterization  of the  faintest
population  of  halo  white  dwarfs.    All  this,  in  turn,  reduces
considerably the  size of the  observational sample, since  at present
large volumes  cannot be probed,  and we  are limited to  study nearby
halo white dwarfs. 

Nevertheless, recent  observational attempts to  empirically determine
the luminosity function  of halo white dwarfs have come  to a success,
and   we  now   have  a   reliable   sample  of   halo  white   dwarfs
\citep{Harris2006, Rowell2011}, to which  the theoretical works can be
compared. Comparing  the results  of the  theoretical models  with the
available observed sample  of halo white dwarfs is  an important task,
as we  now also  have accurate  white dwarf  cooling tracks  for white
dwarfs    descending    from    very    low-metallicity    progenitors
\citep{Miller-Bertolami2013, Camissasa}  that improve upon  those used
in  the  early and  pioneering  calculations  of \cite{Isern1998}  and
\cite{Garcia-Berro2004},   and    in   the   more   recent    one   of
\cite{vanOirschot2014}.   These   evolutionary  sequences   have  been
evolved self-consistently from the zero age main sequence, through the
red giant and thermally pulsing AGB  phases to the white dwarf regime,
and have revealed  the important role of residual  hydrogen burning in
the  atmospheres of  low-mass white  dwarfs, a  physical process  that
needs verification.  Finally, these kind  of works are also of crucial
importance to pave the road to  future studies of the large population
of halo white  dwarfs that the European astrometric  mission Gaia will
unveil in the next years \citep{Torres2005}.
 
Our paper  aims at  producing synthetic samples  of the  population of
halo  white  dwarfs using  the  most  up-to-date physical  inputs  and
prescriptions  for the  Galactic spheroid  and compare  them with  the
current  observational   data.   It  is  organized   as  follows.   In
Sect.~\ref{sec:code} we briefly describe  the numerical tools employed
in this  work.  It  is followed  by Sect.~\ref{sec:results},  where we
first discuss the effects of residual hydrogen burning, of the adopted
initial mass function, of the assumed density profile for the Galactic
halo, of a population of unresolved binary white dwarfs, and thosee of
the star  formation history.  Finally,  in Sect.~\ref{sec:conclusions}
we summarize our calculations and we draw our conclusions.

\section{The population synthesis code}

\subsection{A brief description of the numerical set-up}
\label{sec:code}

As   in  our   previous  works   \citep{Garcia-Berro1999,  Torres2002,
Garcia-Berro2004,  Torres2005, Cojocaru2014},  we  use  a Monte  Carlo
population  synthesis code,  in this  case adapted  to model  the halo
population.  In the following we describe the most important inputs of
our standard model.

We initially produced a large number of synthetic main sequence stars,
located according to a isothermal sphere density model:

\begin{equation}
\rho(r)\propto\frac{a^2+R_{\sun}^2}{a^2+r^2}
\end{equation}

\noindent   where   $a\approx  5$~kpc   is   the   core  radius,   and
$R_{\sun}=8.5$~kpc  is  the galactocentric  distance  of  the Sun.  We
assigned to each  synthetic star a value for the  mass at the zero-age
main sequence  randomly generated  from the  initial mass  function of
\cite{Kroupa2001}, and  a time  of birth,  randomly assigned  within a
burst of constant star formation  lasting for $1$~Gyr.  The velocities
of simulated halo stars were  randomly drawn from normal distributions
\citep{BinneyTremaine87}:

\begin{equation}
f(v_r,v_t)=\frac{1}{(2\pi)^{3/2}}\frac{1}{\sigma_r\sigma_t^2}
\exp\left[-\frac{1}{2}\left(\frac{v_r^2}{\sigma_t^2}+\frac{v_{t}^2}
{\sigma_t^2}\right)\right]
\end{equation}

\noindent  where  $\sigma_r$ and  $\sigma_t$  --  the radial  and  the
tangential velocity  dispersions, respectively  -- are related  by the
following expression:

\begin{equation}                                 
\sigma_t^2=\frac{V_{\rm c}^2}{2}+\left[1-\frac{r^2}{a^2+r^2}\right]
\sigma_r^2+\frac{r}{2}\frac{{\rm d}(\sigma_r^2)}{{\rm d}r}
\end{equation}

\noindent  which,  to  a  first  approximation,  leads  to  $\sigma_r=
\sigma_t=   {V_{\rm   c}}/{\sqrt{2}}$.    The   velocity   dispersions
$\sigma_r$     and    $\sigma_t$     are    those     determined    by
\citet{Markovic1997}. For the calculations  reported here we adopted a
circular  velocity $V_{\rm  c}= 220$~km/s.   From these  velocities we
obtained the heliocentric velocities of  each simulated star by adding
the   velocity  of   the  local   standard  of   rest  (LSR)   $v_{\rm
LSR}=-220$~km/s, and the peculiar velocity of the Sun.

Next we computed the main  sequence lifetime for each progenitor star,
adopting a set of  evolutionary sequences with metallicity $Z=0.0001$,
which  together with  the  age of  the population  (for  which in  our
reference model we  adopted 14~Gyr), and the  progenitor mass, allowed
us to  determine which stars have  had time to become  white dwarfs at
present time.  We  then obtained the corresponding  masses and cooling
ages for each simulated white dwarf.  It is worth mentioning here that
all the evolutionary sequences -- that is, the progenitor evolutionary
sequences and the white dwarf cooling ones -- employed in our work are
those of \cite{Camissasa}, which were obtained from fully evolutionary
calculations,    and    expand    the   previous    calculations    of
\citet{Miller-Bertolami2013}. Hence, the  main sequence lifetimes, the
relationship linking  the progenitor and  the white dwarf  masses, and
the  cooling   ages  are  all  self-consistently   computed  using  an
homogenous   evolutionary   framework.    This  represents   a   clear
improvement  over the  most recent  calculations of  this kind,  as we
employed self-consistent evolutionary models  of the right metallicity
that incorporate  state-of-the-art prescriptions for all  the relevant
physical processes. Our calculations incorporate a fraction of 20\% of
non-DA white dwarfs, for which we employ theoretical cooling sequences
for white  dwarfs with  pure helium atmospheres.  We elaborate  on the
cooling tracks employed here in Sect.~\ref{sec:cooling-tracks}.  Using
these values we derived the stellar  parameters of each white dwarf in
the synthetic  sample.  Namely, we computed  its luminosity, effective
temperature,  surface   gravity  and   magnitudes  in   the  different
passbands.   A standard  model of  Galactic absorption  was also  used
\citep{Hakkila1997} to obtain reliable apparent magnitudes.

Our synthetic  white dwarf sample is  then passed through a  series of
filters which mimic the selection criteria employed to observationally
select  halo  white  dwarfs  in  a real  sample.   These  filters  are
described in  detail in Sect.~\ref{sec:filters}. After  this procedure
is  followed the  white  dwarf luminosity  function  can be  computed,
except  for  a  normalization  factor.   We  chose  to  normalize  the
theoretical  results to  the density  of white  dwarfs in  the highest
density bin  with finite  error bars  of the  observational luminosity
function,  $M_{\rm  bol}=15.75$.  This  is,  in  fact,  equivalent  to
normalize  the luminosity  function to  the total  population density,
given  that this  bin practically  dominates the  stellar counts.   We
remark at  this point  that in  our fiducial  model only  single white
dwarfs  were  considered,  however  we also  explored  models  with  a
fraction of unresolved binaries in  our calculations. It is also worth
mentioning  here  that  our  simulations include  as  well  a  careful
exploration  of the  effects of  other inputs,  which will  be further
explained in Sect.~\ref{sec:results}.

\subsection{Cooling tracks}
\label{sec:cooling-tracks}

White dwarf  progenitors in the  Galactic halo are characterized  by a
significantly  low  metallicity.   In  the Solar  vicinity,  the  halo
metallicity  distribution  function  peaks at  [Fe/H]$\sim  -1.5$~dex.
Actually,  \cite{Frebel2013}  and  \cite{Carollo2010} found  that  the
Galactic  halo  has  a  dual  population.  The  first  of  these  halo
populations  peaks at  [Fe/H]$\sim -1.6$~dex,  whereas the  second one
peaks  at [Fe/H]$\sim  -2.2$~dex.  All  in all,  it is  clear that  to
adequately  capture  the  essential   properties  of  this  metal-poor
population,  a  set   of  cooling  sequences  of   white  dwarfs  with
hydrogen-rich atmospheres descending  from low-metallicity progenitors
is needed.

\begin{table}[t]
\caption{Number of  synthetic white dwarfs that  survive the different
  observational  cuts for  a typical  Monte Carlo  realization of  our
  standard model.
\label{table:cuts}}
\begin{center}
\begin{tabular}{lcc}
\hline
\hline
Filter                               &  $N_\mathrm{WD}$ & \% \\
\hline
Initial  sample                      &  592 199 & 100    \\
$\mu_{\rm min}$ cut                  &    8 952 & 1.5    \\
$12 < r_{\rm 59F} < 19.75$           &      111 & 0.02  \\
RPMD cut                             &      111 & 0.02  \\
$V_{\rm tan} > 200~{\rm km~s^{-1}}$  &       77 & 0.01   \\
\hline
\end{tabular}
\end{center}
\end{table}

We  interpolate  the  cooling  times using  the  set  of  evolutionary
sequences  of \cite{Camissasa}.   These  cooling  sequences have  been
computed considering stable, residual  hydrogen shell burning in white
dwarf atmospheres  during the  white dwarf  stage, although  they also
provide a  set of cooling tracks  in which this physical  mechanism is
disregarded.      This     is     an    important     issue,     since
\citet{Miller-Bertolami2013}  showed  that   although  in  most  cases
residual hydrogen burning  is not a significant source  of energy, for
white  dwarfs with  hydrogen atmospheres  descending from  progenitors
with very low  metallicity it can become a dominant  source of energy,
and can delay  significantly white dwarf cooling. This  effect is more
noticeable for  low-mass white  dwarfs with luminosities  ranging from
$\log(L/L_{\sun})=-2$  to $-4$.  As  mentioned, we  consider that  the
adoption of this set of  sequences represents a clear improvement with
respect to the most recent calculation of the luminosity of halo white
dwarfs \citep{vanOirschot2014}, which  employed evolutionary sequences
for  progenitors  of  Solar  metallicity. 

Although   the   evolutionary   sequences  for   white   dwarfs   with
hydrogen-rich atmospheres adopted in this work are a clear improvement
over previous attempts to model  the population of single white dwarfs
in the Galactic  halo, a cautionary remark is in  order here. There is
solid  evidence that  old stellar  systems exhibit  an enhancement  of
$\alpha$ elements  \citep{AG60, W62}.   While such an  enhancement has
virtually no effects  on the evolutionary timescales  of initially low
mass stars, they can play a role in the evolution of intermediate mass
stars.  In particular, the resulting  total metallicity is larger than
the one  obtained by assuming  a Solar-scaled composition and,  due to
the increase of the oxygen  abundance, the global abundance of carbon,
nitrogen  and oxygen  is  larger than  the corresponding  Solar-scaled
abundance. This, in turn, has an effect on the evolutionary timescales
of  the progenitor  stars of  typical white  dwarfs.  A  more rigorous
treatment  of low-metallicity  stars should  require the  inclusion of
$\alpha$-enhanced initial chemical compositions to compute the stellar
sequences.  Our evolutionary  sequences do not take  into account this
enhancement, but we  estimate that the effects of including  it in the
calculation of  the white  dwarf luminosity  function is  limited.  In
particular, for  the metallicities and progenitor  masses relevant for
our  study, we  have checked  that the  differences of  the progenitor
lifetimes and of  the resulting white dwarf masses  between the values
obtained with  solar-scaled metallicities  and those obtained  with an
$\alpha$-enhanced   chemical   composition   are  smaller   than   1\%
\citep{Pietrinferni}.  Hence,  our results  are almost  insensitive to
the adopted metal ratios.

Finally, for  more massive  oxygen-neon white  dwarfs we  employed the
cooling sequences  of \cite{One2}  and \cite{ONe1}, whereas  for white
dwarfs  with pure  helium atmospheres  we used  the cooling  tracks of
\cite{Bergeron2011}.   In  both  cases the  white  dwarf  evolutionary
sequences  correspond to  progenitors of  Solar metallicity.  This, of
course, is not  a self-consitent treatment, but  nevertheless we judge
that  the effects  on the  computed white  dwarf luminosity  functions
should be modest, see below for a detailed discussion.

\subsection{The observational sample and its observational cuts}
\label{sec:filters}

We compare our results with the most recent and statistically relevant
observational halo white dwarf luminosity function \citep{Rowell2011}.
This observational luminosity function was derived from a sample of 93
halo white dwarfs  detected in the SuperCosmos Sky  Survey (SSS).  The
SSS is an advanced photographic plate-digitizing machine, using plates
taken with the UK Schmidt telescope (UKST), the ESO Schmidt telescope,
and  the  Palomar  Schmidt  telescope.  The  resulting  catalogs  were
compiled  by digitizing  several generations  of photographic  Schmidt
plates. The survey uses a photometric system that has three passbands:
$b_{\rm      J}$,     $r_{\rm      59F}$,     and      $i_{\rm     N}$
\citep{Hambly2001}. Employing data from several generations of plates,
\cite{Rowell2011} constructed a catalog  of $\sim$~10,000 white dwarfs
with magnitudes down  to $r_{\rm 59F} \sim 19.75$,  and proper motions
as  low  as  $\mu  \sim 0.05~{\rm  yr^{-1}}$,  covering  nearly  three
quarters of the sky. Using  strict velocity cuts, the authors isolated
subsamples of white dwarfs belonging to  the thin disk, the thick disk
and  the halo  populations,  and presented  observational white  dwarf
luminosity functions for each one of these populations.

In our study  we distinguish between the complete  sample of synthetic
halo white dwarfs  and a restricted sample. The latter  is obtained by
replicating the observational selection criteria adopted to derive the
observed halo  white dwarf luminosity  function of the SSS.   First, a
proper  motion  cut   depending  on  the  $b_{\rm   J}$  magnitude  is
applied. This proper motion cut  is given by the following expression:
$\mu    >   5(\sigma^{\rm    max}_{\mu}(b_{\rm   J})+0.002)$,    where
$\sigma_{\mu}$  is  the  standard   deviation  in  the  proper  motion
measurements. Also  a magnitude cut  is imposed,  $12 < r_{\rm  59F} <
19.75$.   Next,  a  cut  in  the  reduced  proper  motion  diagram  is
performed,  selecting only  objects below  and blueward  of a  reduced
proper  motion corresponding  to $V_{\rm  tan} =  30~{\rm km~s^{-1}}$.
Lastly,  in  order  to  separate the  halo  population,  a  tangential
velocity  cut  is  used.  Specifically,  we  only  select  stars  with
tangential velocities  $V_{\rm tan} > 200~{\rm  km~s^{-1}}$.  Finally,
we also impose an upper limit  on the tangential velocity of $400~{\rm
km~s^{-1}}$ to prevent selecting stars with velocities larger than the
escape velocity of the Galaxy.

\section{Results}
\label{sec:results}

\begin{figure}[t]
   \centering 
   \includegraphics[width=0.95\columnwidth]{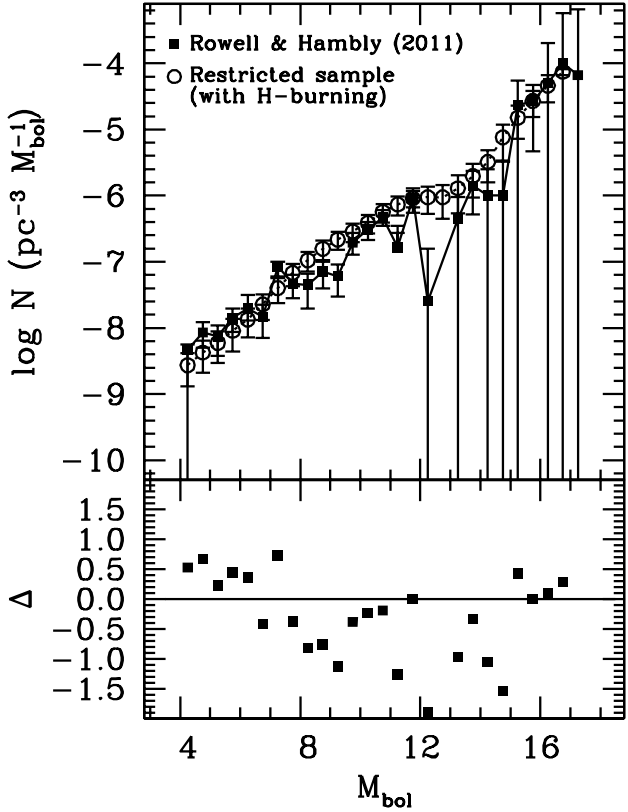}
   \caption{Halo  white dwarf  luminosity  function  for our  fiducial
     Galactic model.  The top panel  shows the theoretical white dwarf
     luminosity   function  obtained   when   the  cooling   sequences
     incorporating  residual  hydrogen   burning  are  employed  (open
     circles). We also  show the observed halo  luminosity function of
     \citet{Rowell2011} -- solid squares.   The bottom panel shows the
     residuals  between  the  observed luminosity  function,  and  the
     theoretical   calculations,   $\Delta  =   2(N_{\rm   obs}-N_{\rm
     sim})/(N_{\rm obs}+N_{\rm sim})$.}
\label{fig:burning}
\end{figure}

In this section we compare the  results of our simulations to the halo
luminosity   function  of   \citet{Rowell2011},  and   we  study   the
sensitivity  of the  theoretical  white dwarf  luminosity function  to
different model inputs.

\begin{figure}[t]
   \centering
   \includegraphics[width=0.95\columnwidth]{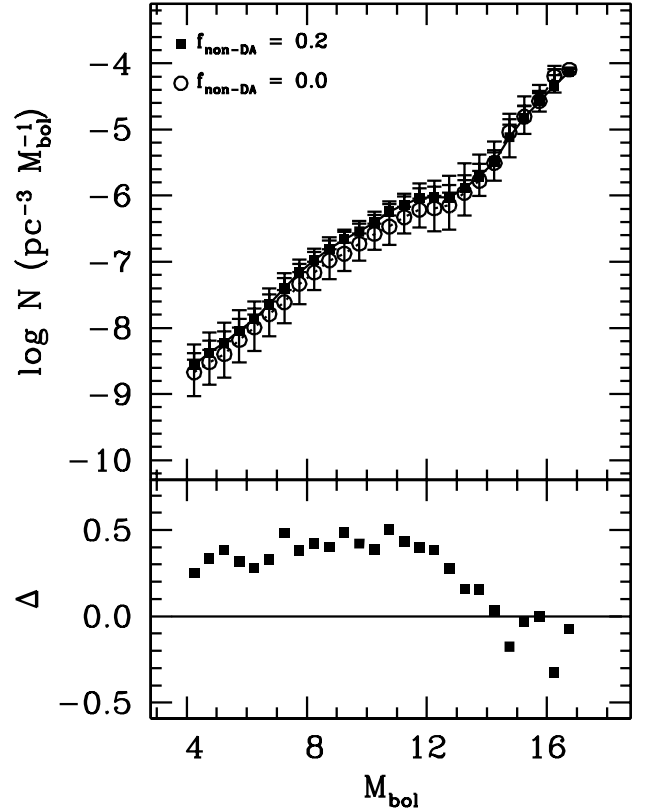}
   \caption{White dwarf  luminosity functions when  only hydrogen-rich
     synthetic  white   dwarfs  are  generated  in   the  Monte  Carlo
     simulation.   The bottom  panel shows  the residuals  between our
     standard  model, in  which a  fraction  of 20\%  of non-DA  white
     dwarfs was employed,  and that in which this  percentage is zero,
     $\Delta  =   2(N_{\rm  std}-N_{\rm   no-DA})/(N_{\rm  std}+N_{\rm
     no-DA})$ --- solid squares and hollow circles, respectively.}
   \label{fig:non-DA}
\end{figure}

To start  with, we  discuss how  the observational  selection criteria
affect the size  of the synthetic samples. This is  done with the help
of Table~\ref{table:cuts}.   In this table  we list for  our reference
model  the number  of white  dwarfs of  the original  synthetic sample
(first row) and in subsequent rows  we list the number of white dwarfs
that survive  the different cuts.   As can be  seen only 1.5\%  of the
synthetic stars  survive the  proper motion  cut.  After  applying the
magnitude cut we are left with 111 synthetic stars, representing about
0.02\% of  the original sample.   For this particular  realization the
reduced  proper motion  cut does  not decrease  further the  number of
simulated white  dwarfs, whereas  the filter in  tangential velocities
reduces even more the number of simulated stars to about 0.01\% of the
original sample,  to 77  white dwarfs, a  number comparable  with that
observationally found.

\begin{figure}[t]
   \centering
   \includegraphics[width=0.95\columnwidth]{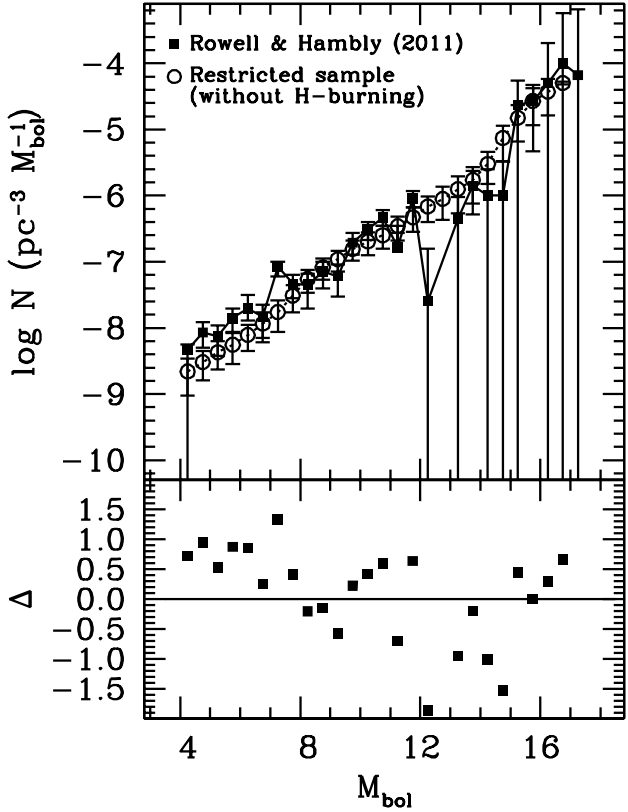}
   \caption{Same  as  Fig.~\ref{fig:burning}  for the  case  in  which
     residual  hydrogen burning  is not  considered. The  bottom panel
     shows  the residuals  between  the  luminosity function  computed
     using  our  standard cooling  sequences  and  that obtained  when
     residual  hydrogen  burning  is artifically  ignored,  $\Delta  =
     2(N_{\rm std}-N_{\rm no-H})/(N_{\rm std}+N_{\rm no-H})$.}
\label{fig:noburning}
\end{figure}

To compare  our results with  those of \cite{vanOirschot2014}  we only
culled  white  dwarfs  using  the tangential  velocity  cut,  as  they
did. Using  only this selection  criterion the size of  the restricted
sample is much  larger. In particular, when this  procedure is adopted
it results  in a restricted sample  which is 63\% of  the initial one.
Obviously, the advantage  of such a large synthetic sample  is that it
is  comparable to  the  complete one,  producing  a smooth  luminosity
function  that faithfully  preserves  the intricacies  of the  adopted
model.  However,  the main  drawback of  adopting just  this selection
criterion  is that  the resulting  sample is  ultimately not  directly
comparable to the observational one, as is ours.

\begin{figure}[t]
   \centering
   \includegraphics[width=0.95\columnwidth]{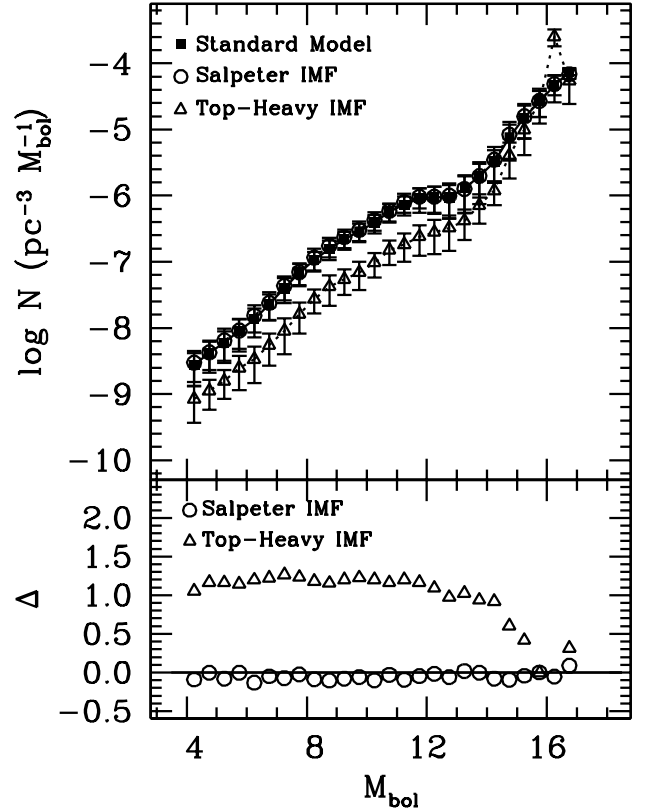}
   \caption{White  dwarf luminosity  functions when  different initial
     mass  functions  are considered  for  the  stellar spheroid.  The
     bottom panel shows  the residuals between our  standard model and
     those  obtained when  the \cite{Salpeter1955}  and the  top-heavy
     initial  mass  function of  \cite{Suda2013}  are used, $\Delta  =
     2(N_{\rm  std}-N_{\rm IMF})/(N_{\rm  std}+N_{\rm IMF})$  --- void
     circles and triangles, respectively.}
   \label{fig:IMF}
\end{figure}

The  top  panel  of   Fig.~\ref{fig:burning}  shows  the  white  dwarf
luminosity function of our reference model  -- open circles -- and the
observed luminosity function of  \cite{Rowell2011} -- solid squares --
while in the  bottom panel the corresponding  residuals are displayed.
As can be seen, the agreement  between the theoretical results and the
observed data is very good. Our fiducial model reproduces not only the
observed  slope  of the  white  dwarf  luminosity function,  but  also
accounts  for  the   scarcity  of  halo  white  dwarfs   at  very  low
luminosities ($M_{\rm bol}>17$).  This  indicates that our Monte Carlo
code  correctly   reproduces  the   selection  criteria   employed  by
\cite{Rowell2011}.

In a  second step, we  have checked  the sensitivity of  our synthetic
white dwarf luminosity function to our choice of cooling sequences for
massive, ONe white dwarfs, and non-DA white dwarfs. We recall that for
these  stars  we  employed  a   set  of  cooling  sequences  of  Solar
metallicity.  Specifically, we  assessed which is the  final number of
these  white dwarfs  in the  restricted sample  -- that  is, once  the
observational  selection criteria  are taken  into account  -- and  we
found that  in a  typical Monte  Carlo realization  only one  of these
white dwarfs,  at most, survives  the successive selection  cuts.  The
most stringent  observational cut is  the magnitude cut,  $r_{\rm 59F}
\sim  19.75$. In  most Monte  Carlo realizations  none of  these white
dwarfs survives this  cut. Additionally, we mention that  even if this
cut is not  employed, the proper motion cut eliminates  from the final
sample almost 99.5\% of ONe white dwarfs. Thus, there are very few ONe
white dwarfs  in the final  sample.  The  reason for this  behavior is
twofold.   First, these  white  dwarfs are  very  scarce, since  their
formation  is strongly  inhibited by  the  shape of  the initial  mass
function. Thus, not surpringsily, they  contribute little to the white
dwarf luminosity function. The second reason is that being these white
dwarfs made  of oxygen and neon,  their heat capacity is  smaller than
that of a  carbon-oxygen white dwarf of the  same mass \citep{ONeold},
and  consequently  cool  faster.    Accordingly,  these  white  dwarfs
contribute  essentially  to  the   faintest  bins  of  the  luminosity
function, which is not probed  by observations, because it is excluded
by the  magnitude cut. In summary,  we conclude that the  influence of
adopting a set of cooling sequences of Solar metallicity for ONe white
dwarfs is negligible.

To assess  the influence  of adopting  a set  of cooling  sequences of
Solar  metallicity  for  non-DA  white dwarfs  we  ran  an  additional
simulation in which  the percentage of non-DA white dwarfs  was set to
zero,  and  consequently all  the  synthetic  stars had  hydrogen-rich
atmospheres.   We then  computed the  residuals between  the resulting
white dwarf luminosity  function and that obtained  with our reference
model, for  which this percentage is  20\%.  The results are  shown in
Fig.~\ref{fig:non-DA}.   As can  be  seen the  differences are  small,
although not negligible. As a matter of fact, the space density of hot
white dwarfs is  smaller in the case in which  only synthetic DA white
dwarfs  are  generated.   However,  this   is  a  consequence  of  the
normalization procedure. Hydrogen-defficient white dwarfs have cooling
sequences that resemble those of a black body, whereas the atmospheres
of  DA  white  dwarfs  are more  transparent.   Consequently,  at  low
temperatures non-DA  white dwarfs  cool faster than  DA ones,  and the
percentage   of  non-DA   white   dwarfs   increases  for   decreasing
luminosities, so  these white dwarfs accumulate  at luminosities close
to that  of the peak of  the theoretical luminosity function  and even
smaller.  However, the number counts of white dwarfs in the luminosity
bins close to  the peak of the luminosity function  dominate the total
number counts of white dwarfs in the synthetic sample. Thus, since the
total number  of white dwarfs in  any Monte Carlo realization  must be
kept constant and, moreover, must be  close to the observed value, the
hot branch of the luminosity function is depleted in the case in which
non-DA  white dwarfs  are not  generated.  Nevertheless,  we emphasize
that  because white  dwarf cooling  sequences of  low metallicity  for
non-DA white dwarfs are not available, it is clear that this procedure
largely  overestimates  the  impact  of  adopting  a  set  of  cooling
sequences of  unappropriate metallicity.   Thus, we conclude  that the
possible  effect of  adopting  a  set of  cooling  sequences of  Solar
metallicity for non-DA white dwarfs is limited.

Next, we assess the sensitivity of  these results to the most relevant
inputs of  our model. In particular,  we first discuss if  the adopted
cooling  tracks  for  carbon-oxygen white  dwarfs  with  hydrogen-rich
atmospheres  could change  this picture.  In a  second step,  we study
whether a different choice of  the adopted initial mass function could
affect our results. Later, we evaluate if a different halo model could
have a  noticeable influence  in our  calculations.  Finally,  we also
study if different  percentages of unresolved binaries  vary the shape
of  the  white  dwarf  luminosity function.   We  end  our  assessment
comparing our  theoretical resuls  for different  ages of  the stellar
halo.

\subsection{Hydrogen burning}

\begin{figure}[t]
   \centering
   \includegraphics[width=0.95\columnwidth]{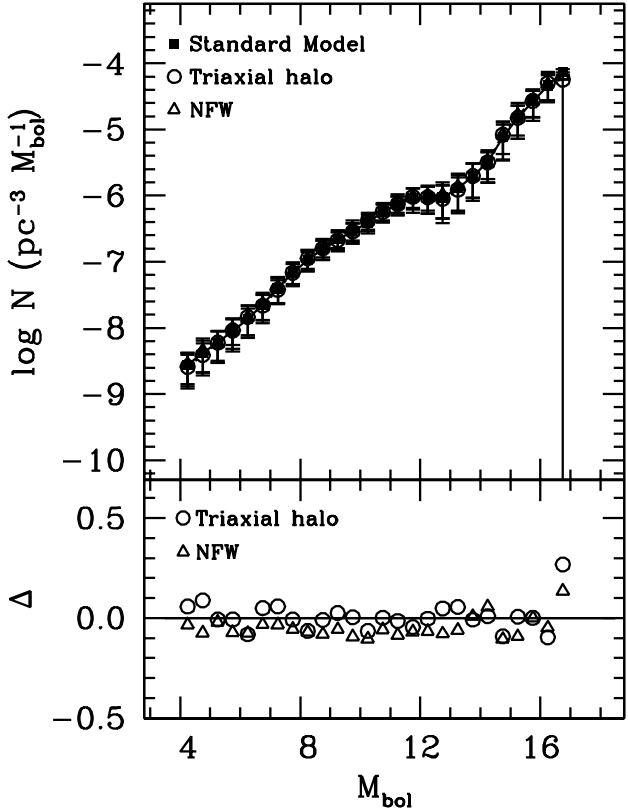}
   \caption{White  dwarf luminosity  functions  for different  density
     profiles  of  the  stellar  halo. The  bottom  panel  shows  the
     residuals between our standard model  and those obtained when the
     density profile of a triaxial  halo \citep{Helmi2004} and that of
     \cite{Navarro1996}    are    employed,   $\Delta    =    2(N_{\rm
     std}-N_{\rho})/(N_{\rm  std}+N_{\rho})$  --- void  circles  and
     triangles, respectively.}
   \label{fig:profiles}
\end{figure}

It has been shown  \citep{Miller-Bertolami2013} that residual hydrogen
burning can have a significant impact  on the cooling process of white
dwarfs with progenitors of very low metallicity, the effect being more
noticeable for low-mass white dwarfs  (those with masses between $0.5$
and $0.6\, M_{\sun}$).  Since low-mass  white dwarfs contribute to all
the luminosity bins of the hot  branch of the luminosity function, and
since the shape of the luminosity  function is directly related to the
cooling  rate, it  is natural  to  ask ourselves  whether a  different
choice of cooling sequences could  affect the slope at moderately high
luminosities.  We check  this using the two different  sets of cooling
tracks described in \cite{Camissasa}.  The  first of these sets is the
one  used  in our  reference  model,  and considers  residual  nuclear
burning,  while the  second one  does  not take  into account  nuclear
reactions (as it occurs for white dwarf progenitors with $Z>0.001$).
 
In  Fig.~\ref{fig:noburning}  we  present the  resulting  white  dwarf
luminosity function for the halo population when the cooling sequences
in  which  residual  hydrogen  burning  is  artificially  ignored  are
adopted. This luminosity  function should be compared  with that shown
in Fig.~\ref{fig:burning}. As can be seen, the only difference between
both sets  of theoretical calculations is  that for the case  in which
the  cooling sequences  incorporating  residual  hydrogen burning  are
employed there is a small plateau between $M_{\rm bol} = 12$ and $14$,
which is absent  in the case in which no  residual hydrogen burning is
considered.  This plateau reflects the slow-down of cooling due to the
release  of   energy  of  residual  hydrogen   burning.   However  the
differences  between both  calculations are  minor, and  the currently
available observational  luminosity function  (which is  derived using
only  $\sim  100$  white  dwarfs)  does not  allow  to  draw  definite
conclusions about the real existence of residual nuclear burning.

\subsection{Initial mass function}

\begin{figure}[t]
   \centering
   \includegraphics[width=0.95\columnwidth]{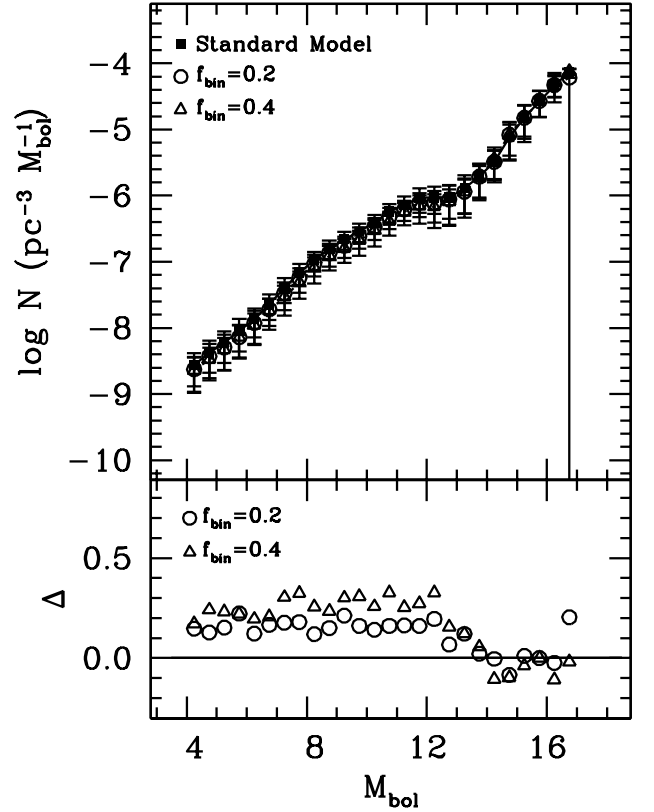}
   \caption{Same   as   Fig.~\ref{fig:profiles}  but   for   different
     fractions  of unresolved  binaries.  The  bottom panel  shows the
     residuals between our standard model  and those obtained when the
     adopted  fractions  of unresolved  binaries are  20\%  and  40\%,
     $\Delta =  2(N_{\rm std}-N_{\rm bin})/(N_{\rm  std}+N_{\rm bin})$
     --- void circles and triangles, respectively.}
   \label{fig:binaries}
\end{figure}

\begin{figure}[t]
   \centering
   \includegraphics[width=0.95\columnwidth]{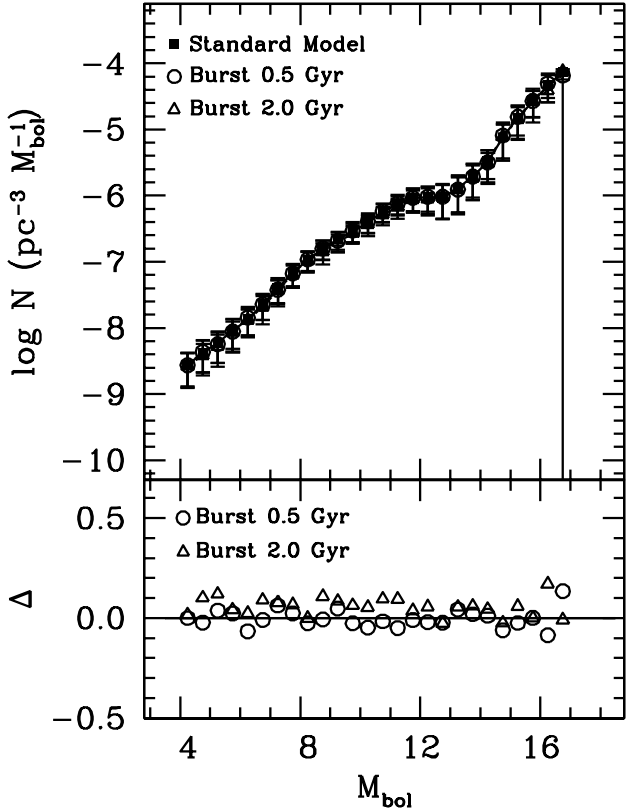}
   \caption{Same   as   Fig.~\ref{fig:profiles}  but   for   different
     durations  of the  initial burst  of star  formation. The  bottom
     panel shows  the residuals between  our standard model  and those
     obtained when the adopted durations  of the initial burst of star
     formation are 0.5 and  2.0~Gyr, $\Delta = 2(N_{\rm std}-N_{\Delta
     t})/(N_{\rm std}+N_{\Delta  t})$ --- void circles  and triangles,
     respectively.}
   \label{fig:burst}
\end{figure}

As mentioned, we also test the influence that the adopted initial mass
function may have in our results. Since the formation timescale of the
stellar halo is short, it is straightforward to show that when a burst
of negligible duration is adopted the luminosity function is given by
\begin{equation}
N(L)\propto\frac{dn}{d M_{\rm bol}}=\frac{dn}{dM}
           \frac{dM}{d M_{\rm bol}}\propto\Phi(M)\frac{dM}{d M_{\rm bol}}
\end{equation}
In this  expression $n$ stands for  the space density, and  $\Phi$ for
the initial mass  function (IMF).  Thus, it is clear  that the adopted
initial mass  function should  influence the  shape of  the luminosity
function.

In order to  test the influence of the IMF  on the luminosity function
we employ  three parametrizations. The first  one is that used  in our
fiducial model,  namely the  so-called ``universal'' mass  function of
\cite{Kroupa2001}.  For the mass ranges relevant to our study this IMF
is  totally  equivalent  to  a  two-branch  power  law  with  exponent
$-\alpha$,  with $\alpha  = 1.3$  for $0.08  \leq M/M_{\sun}<0.5$  and
$\alpha = 2.3$ for $M/M_{\sun}  \geq 0.5$. We also compute theoretical
white  dwarf  luminosity  functions  adopting  the  classical  IMF  of
\cite{Salpeter1955},  which  is  a  power law  with  index  $\alpha  =
2.35$. Finally we also adopt a top-heavy IMF:
\begin{equation}
\Phi(M) = \frac{1}{M} \exp\left(\frac{-\log(M/\mu)}{2\sigma^2}\right)
\end{equation}
In this expression $\mu=10\, M_{\sun}$ and $\sigma=0.44$. This IMF was
introduced by  \cite{Suda2013}, and is  dominated by high  mass stars.
It has been found that  this IMF better reproduces the characteristics
of metal-poor populations, namely those with [Fe/H]$\leq-2$.

The corresponding luminosity functions for these IMFs are shown in the
top panel  of Fig.~\ref{fig:IMF}, and their  respective residuals with
respect to  our fiducial model are  shown in the bottom  panel of this
figure.  As can  be seen, there are no  noticeable differences between
the calculations  in which  the IMF of  \cite{Kroupa2001} and  that of
\cite{Salpeter1955} are employed.  The reason  for this is that in the
relevant luminosity range  the slope of both IMFs is  very similar. We
note, however, that when the  top-heavy IMF of \cite{Suda2013} is used
the luminosity function presents a drop  in the space density at large
luminosities. This deficit  of bright white dwarfs  is quite apparent,
but it is marginally consistent with the observed data.

\subsection{Density profiles}

Another possible concern would be  the adopted density profile for the
stellar halo.  As explained  in Sect.~\ref{sec:code}, in our reference
model  we adopted  the  density profile  of  the classical  isothermal
sphere, but there are other  density profiles that are worth studying.
Accordingly here we  study how this choice affects our  results. To do
this we first adopted a triaxial  oblate halo model, which is based in
a logarithmic dark halo potential \citep{Helmi2004},
\begin{equation}
V=\frac{1}{2}{v_0}^2\ln(R^2+z^2/q^2+d^2)
\end{equation}
being the corresponding density distribution:
\begin{equation}
\rho(R,z)=\left(\frac{v_0^2}{4\pi Gq^2}\right)
\frac{(2q^2+1)d^2+R^2+(2-q^{-2})z^2}{(d^2+R^2+z^2q^{-2})^2}
\end{equation}
In this  expression we  have adopted $d=12$~kpc  and $v_0=131.5$~km/s,
which  gives a  circular  velocity  of the  Sun  of  229~km/s, and  an
oblatness parameter $q=0.8$.
The third and  last profile we used is that  of \citep{Navarro1996}, a
widely used one:
\begin{equation}
\rho \sim \left(\frac{r}{r_{s}}\right)^{-1}\left(1-\frac{r}{r_{s}}\right)^{2}
\end{equation}
with $r_{s}=18~$kpc.

As  Fig.~\ref{fig:profiles}  reveals,   the  differences  between  the
luminosity  functions computed  using  these  three different  density
profiles for the stellar halo  are totally negligible. This is because
the sample of halo white dwarfs of \cite{Rowell2011} is local, whereas
the differences between  the three model profiles  should be prominent
at large distances.

\subsection{Unresolved binaries}

\begin{figure*}[t]
   \centering
   \includegraphics[width=0.95\columnwidth]{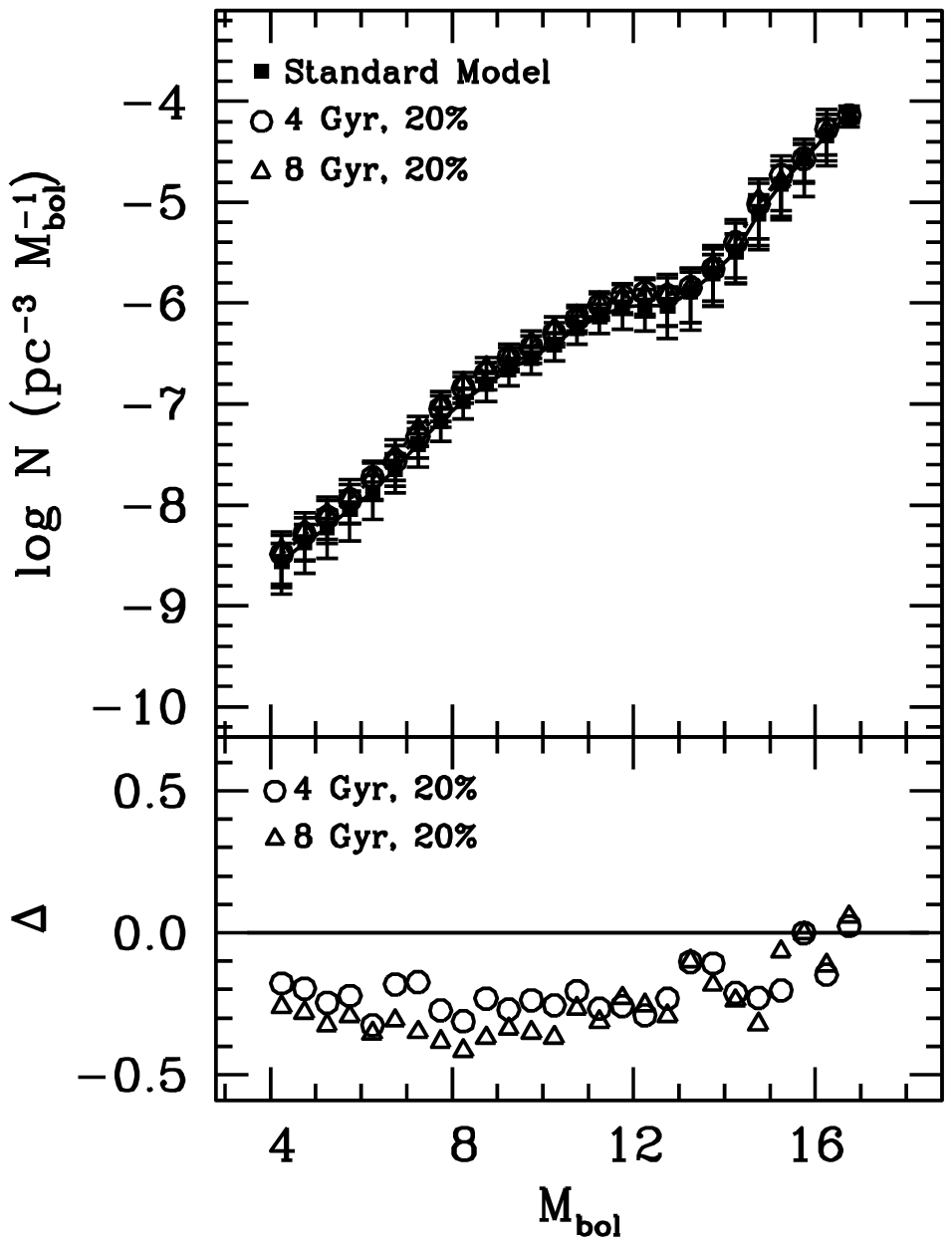}
   \includegraphics[width=0.95\columnwidth]{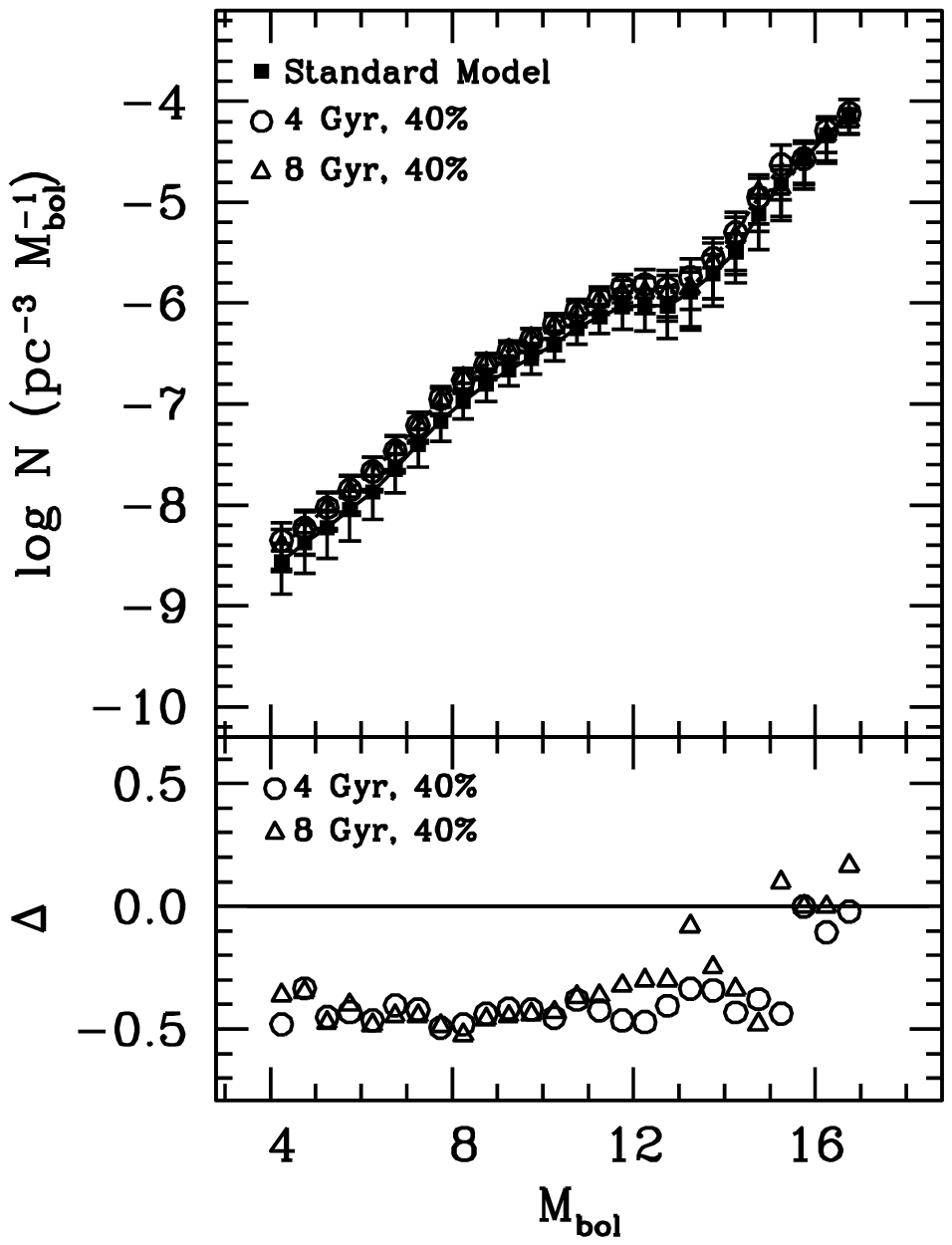}
   \caption{Same  as  Fig.~\ref{fig:profiles}  but  for  four  various
     merger episodes,  of two  strengths at  two different  times. The
     bottom panels show  the residuals between our  standard model and
     those obtained when  the impact of a merger  episode is analyzed,
     $\Delta   =   2(N_{\rm  std}-N_{\rm   mer})/(N_{\rm   std}+N_{\rm
     mer})$. See text for details.}
   \label{fig:accretion}
\end{figure*}

One of the potential problems when calculating the observed luminosity
function for  single stars are  unresolved binary white  dwarfs, since
they compute  as single  stars, and  hence can  result in  a different
shape of the luminosity function. This  has been proven to be the case
in some Galactic clusters  \citep{Bedin2008, Garcia-Berro2010}.  It is
therefore interesting to  check the effect that a  certain fraction of
unresolved binaries  can have on the  theoretical luminosity function.
In order to  test this we compute  a new set of  simulations, based on
our  fiducial  model,  and   increasing  the  fraction  of  unresolved
binaries.   We  remind  that  in our  reference  model  no  unresolved
binaries were considered. As for  the distribution of secondary masses
we  adopted  a model  in  which  the  masses  of both  components  are
uncorrelated.

Fig.~\ref{fig:binaries} shows the result  of this numerical experiment
when the fractions of unresolved  binaries are, respectively, 20\% and
40\%. As  can be observed in  this figure, increasing the  fraction of
unresolved binaries  considered in the  sample does not result  in any
noticeable change,  but in a slight  reduction of the number  of white
dwarfs populating the brightest luminosity  bins.  The reason for this
can  easily  be explained.   Since  low-luminosity  white dwarfs  have
longer evolutionary timescales the low-luminosity bins have also large
space densities. Consequently, unresolved binaries also concentrate in
the luminosity  bins with the  largest densities, and thus  the bright
luminosity bins are less populated. Since we normalize our theoretical
luminosity  function  to  the   observed  luminosity  bin  at  $M_{\rm
bol}=15.75$, the result  is that the bright branch  of the theoretical
luminosity  function is  depleted. Nevertheless,  the differences  are
minor even  when an unrealistic percentage  of 40\% of the  objects in
the synthetic sample are unresolved binaries.

\subsection{The star formation history}

Another point of  concern is the adopted star  formation history. This
also may have potential effects on the morphology of the hot branch of
the halo white dwarf luminosity function. To start with we discuss the
effects of the duration of the  initial burst of star formation.  This
is  done with  the help  of  Fig.~\ref{fig:burst}, where  we show  the
theoretical  white  dwarf  luminosity   functions  for  two  burst  of
durations 0.5 and 2.0~Gyr, and  compare them with our reference model,
for which  we recall  we employed  a burst  of duration  1.0~Gyr. This
figure clearly shows  that, except for the smaller  space densities at
moderately  high  luminosities,  the  differences  between  these  two
luminosity functions and our  fiducial one are marginal. Consequently,
current  observations  do  not  allow  to  discern  between  different
durations of the initial burst of star formation.

\begin{figure}[t]
   \centering
   \includegraphics[width=0.95\columnwidth]{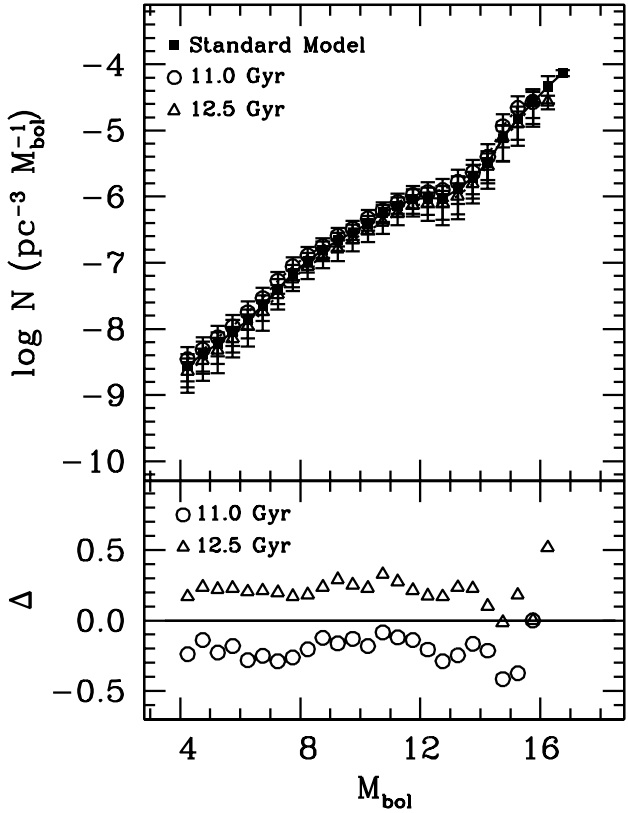}
   \caption{Same as Fig.~\ref{fig:profiles} but  for different ages of
     the halo population. The bottom panel shows the residuals between
     our standard model and the luminosity functions obtained when the
     age  of   the  stellar  halo   is  varied,  $\Delta   =  2(N_{\rm
     std}-N_{T})/(N_{\rm std}+N_{T})$.}
   \label{fig:age}
\end{figure}

Furthermore, a consensus about the  origin of the stellar spheroid has
not been reached  yet. The two main competing  scenarios -- monolithic
collpase  of the  protogalactic  gas  \citep{Eggen1962}, or  formation
through several merger episodes \citep{Searle1978} -- need still to be
confronted with observations. Hence, it  is natural to wonder if these
two scenarios leave observable imprints in the shape of the hot branch
of the  white dwarf luminosity  function of single halo  white dwarfs.
To this end we conducted an additional set of simulations in which, in
addition  to the  initial  burst  of star  formation,  we modeled  the
luminosity function  when a second  burst of star  formation occurring
some  time  ago is  adopted.   Specifically,  we ran  four  additional
simulations in  which a  secondary burst of  star formation  occurs at
times 4 and 8~Gyr respectively, varying the strength of this secondary
burst.  The  metallicities of the  secondary bursts of  star formation
were the same  adopted for the initial one. This  choice minimizes the
effects  of such  merger episodes,  but the  effects of  the different
metallicity  of  the  secondary  bursts  are  expected  to  be  minor.
Specifically, the secondary  burst was given amplitudes  20\% and 40\%
of the  initial one.  In  all cases, the  durations of all  the bursts
(that is, both the initial and the secondary ones) were kept fixed and
equal  to 0.1~Gyr,  while  we  recall that  in  the  standard model  a
duration  of  1~Gyr  was  adopted.   The  results  of  this  numerical
experiment  are displayed  in Fig.~\ref{fig:accretion}.   In the  left
panels of this figure we show  the results when a secondary burst with
an amplitude 20\% of the initial  one is adopted, whereas in the right
panels  the results  when  the  amplitude of  the  secondary burst  is
increased to 40\%  of the primary one are displayed.   As can be seen,
the  differences  are  again  very small.   Thus,  unfortunately,  the
current observational database of halo  white dwarfs does not allow to
distinguish the two aforementioned  formation scenarios of the stellar
halo.

\subsection{Age of the population}

Finally, we ran a set of simulations in which we varied the age of the
halo population,  from 11 to  13~Gyr, and  we compared the  results of
these  calculations with  that obtained  in our  reference model,  for
which we remind we adopted an age  of 13.7~Gyr. We show the results of
these  calculations in  Fig.~\ref{fig:age}.  As  expected, the  bright
branch  of  the  white  dwarf  luminosity  function  does  not  depend
appreciably  on the  adopted age  of the  stellar spheroid.  Moreover,
since the observed luminosity function does not show a cut-off the age
of the halo population cannot be yet computed using the termination of
the cooling  sequence of halo white  dwarfs. This is a  consequence of
the  cuts used  to select  the  observed sample,  and specifically  is
caused  by the  cut in  bolometric magnitude.   The only  quantitative
assessment  about the  age  of the  halo  that can  be  made with  the
available observed data  is to place a lower limit.   This can be done
in a simple way, by imposing that the dimmest populated luminosity bin
of   the  theoretical   white  dwarf   luminosity  function   is  that
observationally found, at $M_{\rm  bol} =17.25$.  Using this procedure
we find that, although it is not possible to fit the halo age, a lower
limit for its age of $12.5$~Gyr can be safely established.

\section{Conclusions}
\label{sec:conclusions}

In this paper we have revisited  the luminosity function of halo white
dwarfs,  in the  light of  the recently  computed white  dwarf cooling
sequences  for low-metallicity  progenitors.  These  cooling sequences
\citep{Miller-Bertolami2013,  Camissasa}  have been  derived  evolving
self-consistently their  progenitors from the zero  age main sequence,
through the  red giant and thermally  pulsing AGB phases to  the white
dwarf regime, and have unveiled  the role of residual hydrogen burning
in the  atmospheres of  low-mass white  dwarfs. In  this sense,  it is
important to  realize that  these evolutionary  calculations superseed
those used in the early and  pioneering calculations of the halo white
dwarf     luminosity     function      of     \cite{Isern1998}     and
\cite{Garcia-Berro2004},    and     in    the    recent     work    of
\cite{vanOirschot2014}.  Moreover, in pursuing  this endeavour we have
employed  a state-of-the-art  numerical code,  which incorporates  not
only the  most recent advances  that allow an accurate  description of
the  Galactic  halo,  but  also   a  detailed  implementation  of  the
observational biases and restrictions,  an issue that most theoretical
calculations  do  not  take  into  account,  thus  impeding  a  sought
comparison with the  observed sample.  This is an  important issue, as
the observed sample of white dwarfs belonging to the Galactic spheroid
suffers from  small statistics.  Moreover,  given that the  density of
halo  white dwars  is  low  and that  this  population  is old,  hence
intrinsically faint, the detection of halo white dwarfs is hampered by
observational  difficulties. Consequently,  the  selection biases  are
important, and we are restricted to compare the theoretical results of
our results with an observational sample plagued with uncertainties.

Since residual hydrogen burning occurs at moderately low luminosities,
say from  $\log(L/L_{\sun})=-2$ to $-4$, the  halo luminosity function
could eventually offer  an unique possibility to  test the reliability
of  these  recent  cooling   sequences.   This  could  have  important
consequences for our understanding of  how white dwarfs are formed and
how their progenitor stars evolve in low-metallicity environments, and
more specifically it  could shed light on the occurrence  of the third
dredge-up  for  metallicities  $\la  10^{-3}$.  We  have  found  that,
unfortunately, the scarcity  of halo white dwarfs  at the luminosities
at which  residual hydrogen burning  occurs prevents us from  making a
meaningful  comparison between  the sequences  which incorporate  this
physical ingredient  and those which  do not.  Thus, this  effort will
have  to  wait  until  we  have  larger  and  more  reliable  samples.
Alternatively, this  could be  done using  the white  dwarf luminosity
functions  of  Galactic  globular  clusters,  of  which  NGC~6397  is,
perhaps, the leading example.

Additionally,  we  have  investigated  whether  or  not  the  observed
luminosity function of  single white dwarfs can be  eventually used to
learn  more about  the stellar  population of  the Galactic  halo.  In
particular, we have studied if it can be used to constrain the initial
mass function of this population,  its star formation history and age,
to probe different  halo density profiles, or possibly  to discern the
fraction  of unresolved  binaries that  may contaminate  observations.
Unfortunately,  our  calculations show  that  the  hot branch  of  the
luminosity function is almost insensitive to all these inputs -- as it
occurs for the disk  white dwarf luminosity function \citep{Isern2008}
-- and consequently that unless we  have a more accurate determination
of  the  luminosity  function  at  large  bolometric  magnitudes  (low
luminosities) there  is no hope  to extract all this  information from
the  observed data.   However, large  space-borne surveys,  like Gaia,
will  provide   us  with   a  large  sample   of  halo   white  dwarfs
\citep{Torres2005}, and hopefully all  this wealth of information will
be extracted in  a near future.  Nonetheless, the  lack of sensitivity
of the hot  branch of the luminosity function of  halo white dwarfs to
all these inputs can be interpreted in a positive way, since it allows
us to obtain a robust statistical measure of the cooling rate of white
dwarfs at low metallicities, and high luminosities.

\begin{acknowledgements}
This research  was partially supported by  MCINN grant AYA2011--23102,
by  the European  Union  FEDER funds,  by the  AGAUR  (Spain), by  the
AGENCIA  through the  Programa  de  Modernizaci\'on Tecnol\'ogica  BID
1728/OCAR,  and  by  the   PIP  112-200801-00940  grant  from  CONICET
(Argentina). RC also acknowledges financial support from the FPI grant
BES-2012-053448.
\end{acknowledgements}


\bibliographystyle{aa}
\bibliography{hLWr}

\end{document}